\documentclass[useAMS]{gGAF2e}

\usepackage{psfrag}
\usepackage{color}

\newcommand{\de}{{\rm \, d}}

\renewcommand{\vec}[1]{\mathbf{#1}}

\newcommand\etal{\mbox{\textit{et al.}}}

\newcommand\red[1]{{#1}}

\allowdisplaybreaks

\begin{document}
\doi{10.1080/03091929.2010.519891}
 \issn{} \issnp{} \jvol{} \jnum{} \jyear{2011}
 \jmonth{}

\markboth{Friedrich H.~Busse and Radostin D.~Simitev}{Remarks on dynamo theory}

\title{Remarks on some typical assumptions in dynamo
  theory$^1$}

\author{FRIEDRICH
  H.~BUSSE${\dag}$$^{\ast}$\thanks{$^\ast$Corresponding author. Email:
    busse@uni-bayreuth.de}
\thanks{$^1$Dedicated to Raymond Hide on the occasion of his 80th birthday.
\vspace{6pt}} 
and RADOSTIN D.~SIMITEV${\ddag}$\\\vspace{6pt}
${\dag}$Institute of Physics, University of Bayreuth, D-95440 Bayreuth, Germany\\
${\ddag}$School of Mathematics and Statistics, University of Glasgow, G12 8QW Glasgow, UK\\
\vspace{6pt}\received{\today} }

\maketitle

\begin{abstract}

Some concepts used in the theory of convection-driven dynamos in
rotating spherical fluid shells are discussed. The analogy between
imposed magnetic fields and those generated by dynamo action is
evaluated and the role of the Elsasser number is considered. Eddy
diffusivities are essential ingredients in numerical dynamo
simulations, but their effects could be misleading. New
aspects of the simultaneous existence of different dynamo states are
described. 
\bigskip

\bigskip
\end{abstract}

\section{Introduction}

Magnetohydrodynamic turbulence, i.e. hydrodynamic turbulence together
with dynamo generated magnetic fields, is a particularly complex
phenomenon that is still far from being fully understood. This is
hardly surprising in view of the fact that the number of degrees of
freedom is roughly doubled in  comparison with its purely hydrodynamic
version. The complexities of the details of MHD-turbulence stand in
stark contrast to the observed simple structures such as planetary
dipolar fields nearly aligned with the axis of rotation or the solar
magnetic cycle with its surprisingly regular period of 22 years. It is
thus understandable that numerous attempts have been made to find \red{simple}
balances or to devise simplifying concepts in order to gain some
understanding of the dynamics of MHD-turbulence. 

In the following the validity of some of the assumptions that
have been widely used in dynamo theory will be discussed. We shall
restrict ourselves to buoyancy-driven motions which are responsible
for the generation of planetary and stellar magnetic fields and even
within this restricted field no completeness will be attempted.

\section{Dynamo generated magnetic fields versus imposed fields}

The simplest approach for analyzing magnetohydrodynamic phenomena is
the consideration of the dynamics of an electrically conducting fluid
in the presence of an imposed magnetic field $\vec B_0$. This approach has been
used by several theoreticians and experimenters \red{in the period
from approximately 1950 to 1970}
 in order to understand the onset of thermal convection in the
presence of a uniform imposed  $\vec B_0$. An account of these
efforts can be found in Chandrasekhar's (1961) famous treatise. In
later years nonlinear aspects of the interaction between convection
and the imposed magnetic field have been studied. We refer to the
review of Proctor and Weiss (1982). 

Of special interest is the role of magnetic fields in the case of
convection in a rotating system. As described in Chandrasekhar's
(1961) book the onset of convection is strongly inhibited in systems
with a vertical axis of rotation or, in a non-rotating system, by an
imposed vertical magnetic field when the fluid is electrically
conducting. The critical value $R_c$ of the  Rayleigh number $R$ for
onset of convection thus increases with the $4/3$-power of the
Coriolis number $\tau$ in the non-magnetic system while it increases
in proportion to the Chandrasekhar number $Q$ in a non-rotating
system. But in systems with both a vertical axis of rotation and an
applied vertical magnetic field the onset of convection is facilitated
by the counteracting effects of the Coriolis force and Lorentz force
such that $R_c$ may increase only linearly with $\tau$ (Eltayeb 1972). The
dimensionless parameters introduced here and the ordinary and magnetic
Prandtl numbers are defined by  
\begin{equation}
R = \frac{\alpha g \Delta Td^3}{\nu \kappa}, 
\qquad 
\tau = \frac{2 \Omega d^2}{\nu}, 
\qquad 
Q=\frac{B^2_0 d^2}{\mu\rho \lambda \nu}, 
\qquad 
P= \frac{\nu}{\kappa}, 
\qquad 
P_m = \frac{\nu}{\lambda},
\label{params}
\end{equation}   
where $B_0, \Omega, \mu, \nu, \kappa, \rho$ and $\lambda$ denote the
flux density of the applied magnetic field, the angular velocity of
rotation, the magnetic permeability, the kinematic viscosity, the
thermal diffusivity, the density and the magnetic diffusivity,
respectively. $\Delta T$ is the applied temperature difference between
the boundaries and $d$ is the distance between them; $\alpha$ is the
coefficient of thermal expansion and $g$ is the acceleration of
gravity. The inverse of $\tau$ is also known as the Ekman number. 

The optimal balance between the Coriolis force and the
Lorentz force which minimizes $R_c$ is determined by
the condition that the  Elsasser number $\Lambda$ which
describes the ratio between the two forces,
\begin{equation}
\Lambda=\frac{B^2_0}{2\Omega \mu\rho \lambda},
\end{equation}   
assumes a value of the order unity  (Chandrasekhar 1961). The
counteracting effect of rotation and magnetic field also operates when
both, the axis of rotation and the imposed magnetic field, are
horizontal, but directed in different direction. This situation is
approximately satisfied in the equatorial region of a rotating
spherical fluid shell. For a zonal magnetic field the onset of
convection in this case has been studied numerically by Fearn
(1979). In the closely related configuration of the rotating annulus
with decreasing height with distance from the axis analytical
expressions for the critical Rayleigh number and the associated
wavenumber can be obtained (Busse 1983). Both, the numerical and the
analytical results agree quite well and exhibit a minimum critical
Rayleigh number for $\Lambda\approx 1$.    

\begin{figure}
\vspace*{3mm}
\begin{center}
\epsfig{file=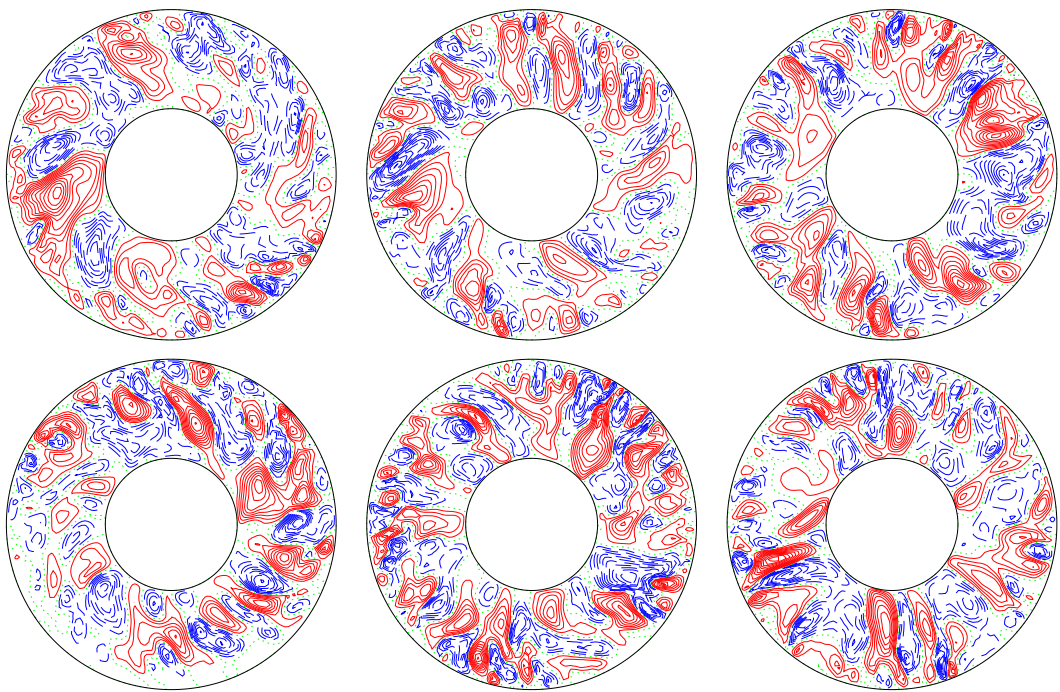,width=4.05cm,width=14cm,clip=}
\end{center}
\caption[]{(Color online)
Comparison of the size of convection rolls
of non-magnetic convection and of self-sustained dynamo solutions.
Snapshots of the equatorial streamlines $r\partial_\varphi v =$ const
at $\tau=3\times10^4$, $R=3.5\times10^6$ in the cases $P=0.5$ (first
row), and $P=0.75$ (second row). The first plot in each row shows the
non-magnetic solutions, while the second and the third plot show the
FD
and MD dynamo with $P_m=1$ (first row), and the FD
and MD dynamo with $P_m=1.5$ (second row), respectively. (See section
4 for description of FD and MD dynamos.)
}
\label{fig-01}
\end{figure}

Because of the special role played by the Elsasser number in
determining the optimal conditions for the onset of convection in the
parameter space spanned by the rate of rotation and the imposed
magnetic flux density, it seemed obvious that the condition
$\Lambda\approx 1$ may determine the strength of the observed
planetary magnetic fields. This idea was  proposed by Eltayeb and
Roberts (1970) and it seems to work fairly well in the cases of
geomagnetism and Jovian magnetism. But the numerous numerical
simulations of convection-driven dynamos in rotating spherical shells
have not supported this hypothesis. First of all, the increase in the
size of convection rolls that is the basic reason for the reduction of
the critical Rayleigh number for $\Lambda\approx 1$ is hardly
noticeable when convection flows with a dynamo field are compared with
flows at the same external conditions, but without a magnetic
field. This has been observed in the work of Christensen {\itshape et al.} (1999)
and is also evident from the streamlines in the equatorial plane shown
for different cases in figure \ref{fig-01}. (The distinction between
MD- and FD-dynamos will be discussed in section 4.) At higher values
of the Coriolis number $\tau$  some small increase in
the scale of convection may be noticed in turning from the
non-magnetic to the dynamo state, but this increase could be attributed
to the higher amplitude of convection which also influences its
characteristic scale. 
\red{All figures presented in this paper are original and have been
  generated from numerical results based on equations \eqref{eq1}
  given in the Appendix. More details on the dynamo simulations can be
  obtained from the papers of Simitev and Busse (2005, 2007).}

\begin{figure}
\psfrag{a}{\sl(a)}
\psfrag{b}{\sl(b)}
\psfrag{0}[tl]{0}
\psfrag{30}[tl]{30}
\psfrag{60}[tl]{60}
\psfrag{90}[tl]{90}
\psfrag{c1}[tl]{\color{red}0}
\psfrag{c2}[tl]{\color{red}30}
\psfrag{c3}[tl]{\color{red}60}
\psfrag{c4}[tl]{\color{red}90}
\psfrag{b1}{\color{blue}0}
\psfrag{b2}{\color{blue}30}
\psfrag{b3}{\color{blue}60}
\psfrag{b4}{\color{blue}90}
\psfrag{-20}{-20}
\psfrag{-80}{-80}
\psfrag{40}{40}
\psfrag{100}{100}
\psfrag{0.8}{\color{red}0.8}
\psfrag{2.7}{\color{red}2.7}
\psfrag{4.6}{\color{red}4.6}
\psfrag{6.5}{\color{red}6.5}
\psfrag{a1}{\color{blue}0.8}
\psfrag{a2}{\color{blue}1.1}
\psfrag{a3}{\color{blue}1.4}
\psfrag{a4}{\color{blue}1.7}
\psfrag{thhhh}[th]{$\theta$ [{\sl deg}]}
\psfrag{thhha}[th]{\color{red}$\theta$ [{\sl deg}]}
\psfrag{uoo}{$\langle \overline{u_\varphi}\rangle_t$}
\psfrag{Nioooo}[r][r]{\color{red}$Nu_i-1$}
\psfrag{Nooooo}{\color{blue}$Nu_o-1$}
\begin{center}
\vspace*{3mm}
\epsfig{file=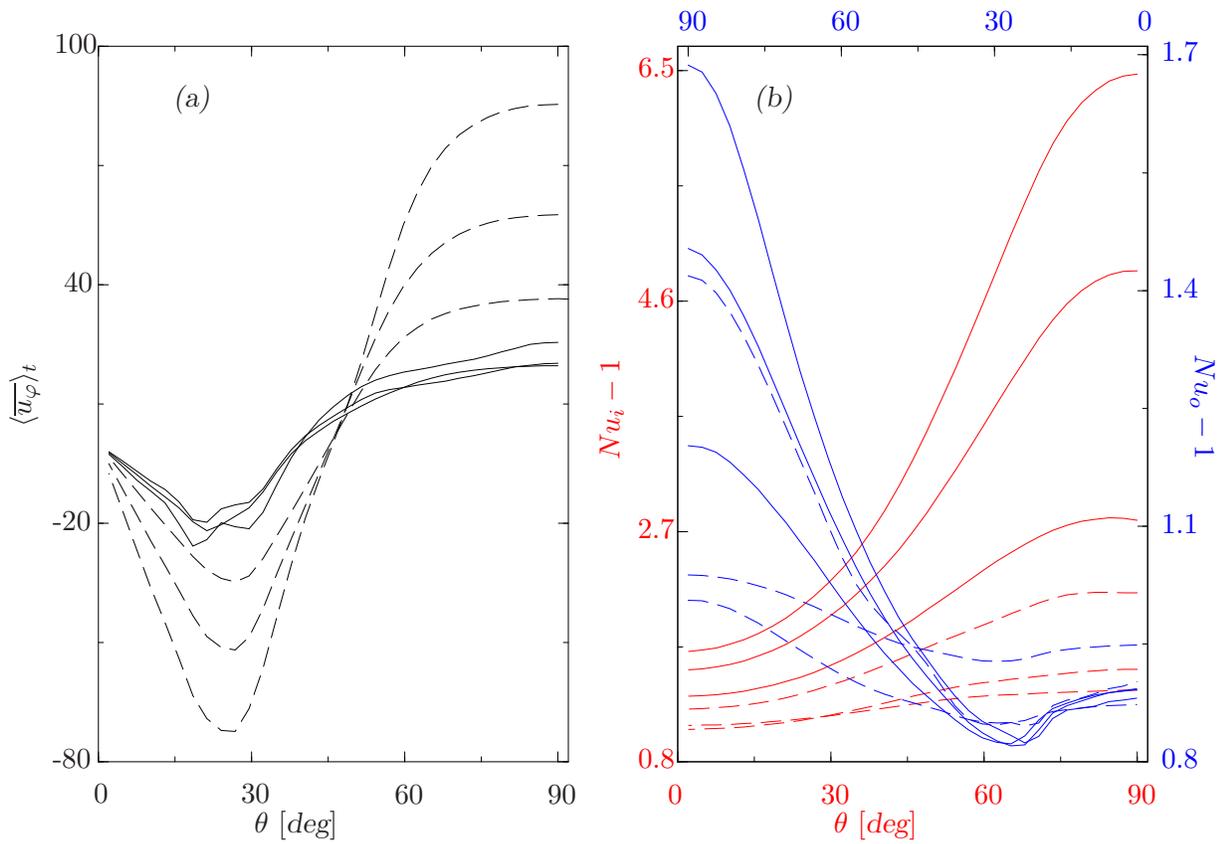,height=16cm,angle=-90,clip=}
\end{center}
\vspace{5mm}
\caption[]{(Color online)
Time- and azimuthally-averaged zonal flow, $\langle
  \overline{u_\varphi}\rangle_t$, at the outer spherical boundary in
  {\sl (a)} and local $Nu_i$ (red lines, left ordinate, lower
  abscissa, curves increasing to the right)  and $Nu_o$ (blue
  lines, right ordinate, upper abscissa, curves increasing to the
  left) 
  in  {\sl (b)} as  functions of the colatitude $\theta$ for $P = 1$,
  $\tau=10^4$. From bottom to top at $\theta=90^\circ$, the
  values of $R$ are $R=5\times10^5$, $6\times10^5$ and
  $7\times10^5$ for both non-magnetic convection (dashed lines) and dynamo
  solutions with $P_m=6$ (solid lines).}
\label{fig-02}
\end{figure}

\red{Both the} imposed magnetic field with $\Lambda\approx 1$ and the
magnetic field generated by dynamo action do  promote convection and
increase the convective heat transport. The dynamo field accomplishes
this mainly by putting the brakes onto the differential rotation
which, although generated by the Reynolds stresses of convection,
exerts an inhibiting influence on the convection eddies through its
shearing action. This
nonlinear process is very different from the effect of a nearly
homogeneous imposed magnetic field. While the latter enters into the
linear terms of the basic equations of motion and thus  reduces the
Rayleigh number for onset of convection, dynamo fields have not shown
such an effect. A typical impression of the effect of the dynamo field
can be gained from figure \ref{fig-02} where the azimuthally averaged
zonal flow and the local  heat transports have been plotted as a
function of the colatitude $\theta$. These plots demonstrate that the dynamo
fields affect primarily the amplitude but not the structure of
convection. For the definitions of the Nusselt numbers $Nu_i$ and
$Nu_o$ see expressions \eqref{nu.def} in the Appendix.  

\section{Eddy diffusivities}

Dissipation is not only an important part of all dynamical processes
such as the dynamo process, but it is also an essential ingredient of
most numerical schemes for the solution of nonlinear partial
differential equations. Difficulties can appear when there exist large
differences in the various diffusivities entering a problem. Examples
are presented by the problem of high Rayleigh number convection at very
low Prandtl numbers when the thermal diffusivity vastly exceeds the
kinematic viscosity of the fluid or in the case of the geodynamo when
it is attempted to approach the molecular values of the magnetic
diffusivity and the kinematic viscosity. 

For the simulation of many turbulent systems the molecular
diffusivities are far too low to be used in the numerical
analysis. Here the concept of eddy diffusivities enters as a
convenient solution. On the one hand, it is supposed to provide a
representation of the diffusive effects of small-scale turbulence that
arises from those components of various fields that can not be
resolved numerically. On the other hand, eddy viscosities are
needed to stabilize numerical schemes. Since small scale turbulence
acts in a similar fashion on many physical quantities such as
momentum, magnetic 
flux and temperature it is usually assumed that all eddy diffusivities
attain the same value and that as a consequence eddy diffusivity
ratios such as the effective ordinary and magnetic Prandtl numbers can
be set equal to unity. 

This procedure can be misleading for several reasons. First of all, the
action of turbulence on a solenoidal vector quantity tends to be
different from the action on a scalar quantity. Secondly, large
differences in the molecular values of diffusivities are not likely to
be entirely eliminated through the effects of turbulence. \red{The Earth's
core offers an example for this problem. The molecular magnetic
diffusivity is so large that its replacement by an even larger eddy
diffusivity would give rise to magnetic Reynolds numbers likely to be below the critical
value for dynamo action.} From the property that the magnetic
diffusivity by far exceeds the kinematic viscosity, it  can not be concluded that
the effective magnetic Prandtl number to be used in numerical
simulations  should be much less than unity.  While usual derivations
from experimental measurements (Gellert and R\"udiger 2008) and
computer simulations (Yousef {\itshape et al.} 2003) yield turbulent magnetic
Prandtl numbers $P_{mt}$ in the range $0.5\lessapprox
P_{mt}\lessapprox 1$, values in excess of unity have been found
(Yousef {\itshape et al.} 2003) when the molecular value of $P_m$ was very
small. In this respect the magnetic Prandtl number behaves in a
similar fashion as the ordinary turbulent Prandtl number as discussed,
for example, in chapter 12 of the book by Kays {\itshape et al.} (2005). In the
case of liquid metals, for instance, which are characterized by a
Prandtl number of the order of $10^{-2}$, values significantly above
unity have been found for the turbulent Prandtl number $P_t$.  

It must be kept in mind that eddy diffusivities are just crude tools
for the simulation of turbulent transport processes and that a fully
satisfactory determination of diffusivity ratios will never be
achieved. 
\red{Considerable efforts have been expended in the development of more advanced concepts of
 large eddy simulations (LED) in extension of methods used in non-magnetic turbulence. We refer to the
work of Matsui and Buffett (2005) and of Matsushima (2005).
But these efforts have been only partially successful according to the discussion by
Roberts (2007). In rapidly rotating systems} the use of anisotropic
eddy diffusivities could lead to a more realistic description of the
effects of small scale turbulence. But no systematic efforts to determine eddy
diffusivity tensors have been made and in view of the doubtful concept
of eddy diffusivity it is questionable whether such efforts would be
worthwhile.  
\begin{figure}
\psfrag{(a)}{(a)}
\psfrag{(b)}{(b)}
\psfrag{E}{$E$}
\psfrag{M}{$M$}
\psfrag{P}{$P$}
\psfrag{1}{1}
\psfrag{a-1}{$10^{-1}$}
\psfrag{a0}{$1$}
\psfrag{a1}{$10$}
\psfrag{a2}{$10^2$}
\psfrag{a3}{$10^3$}
\psfrag{a4}{$10^4$}
\begin{center}
\vspace*{3mm}
\epsfig{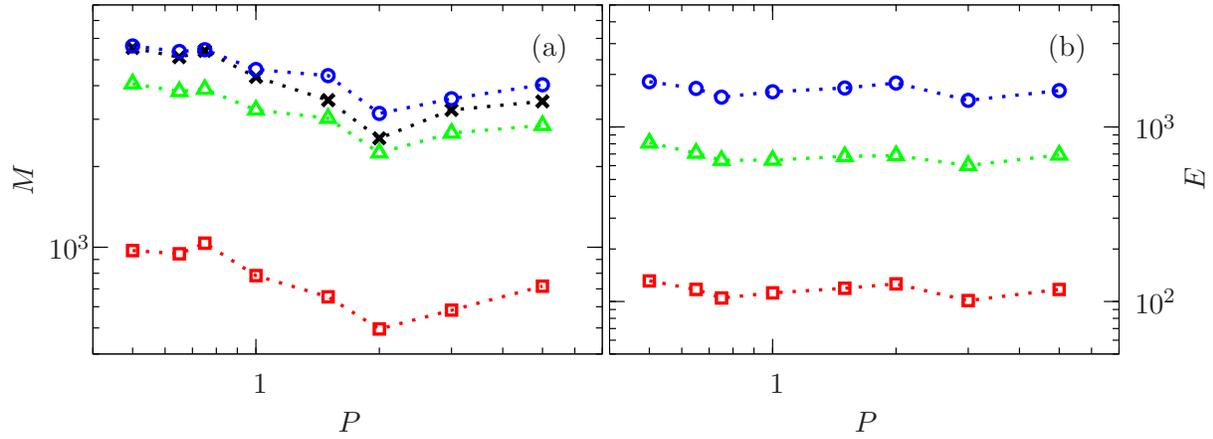}
\end{center}
\caption[]{(Color online) Validity of the magnetostrophic
approximation for MD dynamos at $\tau=3\times10^4$, $R=3.5\times10^6$.
Panel (a) shows the averaged magnetic energy density components
multiplied by $P$ and panel (b) shows the average kinetic energy
density components multiplied by $P^2$, both as a function of $P$ at
fixed $P_m/P=2$.
In both panels, the components $\overline{X}_p$, $\overline{X}_t$,
$\hat X_p$, and $\hat X_t$ are shown by black crosses, red squares,
green triangles and blue circles, respectively, where $X=M,E$.  The
energy density $\overline{E}_p$, has not been included in the right
hand plot since it is nearly two orders of magnitude smaller than
$\overline{E}_t$.
}
\label{fig-02a}
\end{figure}

Finally, it should be noted that diffusivities play different roles in
different situations. Turbulent Ekman layers are not well described by
laminar ones with an eddy 
viscosity replacing the molecular value. Small scale turbulence tends
to cancel the effect of the Coriolis force and to modify Ekman
layers into boundary layers more similar to those found in turbulent
shear flows in non-rotating systems (Coleman {\itshape et al.} 1990). In a
recent paper Miyagoshi {\itshape et al.} (2010) have numerically simulated
convection driven dynamo in rotating spheres with a no-slip outer
boundary at the very low Ekman number of $10^{-7}$. They have come to
the conclusion that the results are more similar to those with a
stress-free outer boundary than to those with a no-slip boundary together
with a more commonly used Ekman number of the order $10^{-5}$ or
larger. We thus tend to conclude that the use of a stress-free
boundary condition may be more realistic in simulations of the
geodynamo than a no-slip boundary with a relatively thick Ekman layer
attached to it. 
  
\section{Parameter dependence of spherical dynamos}

In spite of their shortcomings eddy diffusivities and their ratios will
continue to be used in numerical dynamo simulations for the
foreseeable future. We have already emphasized the importance of
allowing for different values of the various eddy diffusivities. This
importance is amplified by the fact that the properties of turbulent
convection driven dynamos seem to vary most rapidly with $P$ and $P_m$
in the neighborhood where these parameters assume the value unity. 

In the parameter range in which most simulations of convection driven
spherical dynamos have been carried out, two basically different types
of dipolar dynamos have been found. \red{The two types can not be
  distinguished by their  symmetry properties. Only the relative
  amplitudes of various components  of the magnetic and the velocity
  fields differ greatly.} One type which we shall refer to as 
MD-dynamo is characterized by a dominant axisymmetric dipole which in
his ideal manifestation is nearly steady, but which can also oscillate
in time, especially if the Rayleigh number is not very high. The other
type of dynamo referred to as FD-dynamo is dominated by
non-axisymmetric components of the magnetic field with their energy
exceeding the energy of the axisymmetric components. Nevertheless
outside the conducting spherical shell these dynamos may exhibit a
nearly axisymmetric dipole which at lower Prandtl numbers may be
replaced by a hemispherical or an axial quadrupolar field. For even
more dominant non-axisymmetric components the outside field can sometimes be
approximated by an equatorial dipole field. 
\begin{figure}
\begin{center}
\vspace*{3mm}
\psfrag{E}{$\widetilde{E}$}
\psfrag{M}{$\widetilde{M}$}
\psfrag{t}{$t$}
\psfrag{(a)}{(a)}
\psfrag{(b)}{(b)}
\psfrag{0}{0}
\psfrag{1000}{1000}
\psfrag{2000}{2000}
\psfrag{3000}{3000}
\psfrag{4000}{4000}
\psfrag{500} {500} 
\psfrag{1}{1}
\psfrag{2}{2}
\psfrag{3}{3}
\psfrag{4}{4}
\psfrag{5}{5}
\epsfig{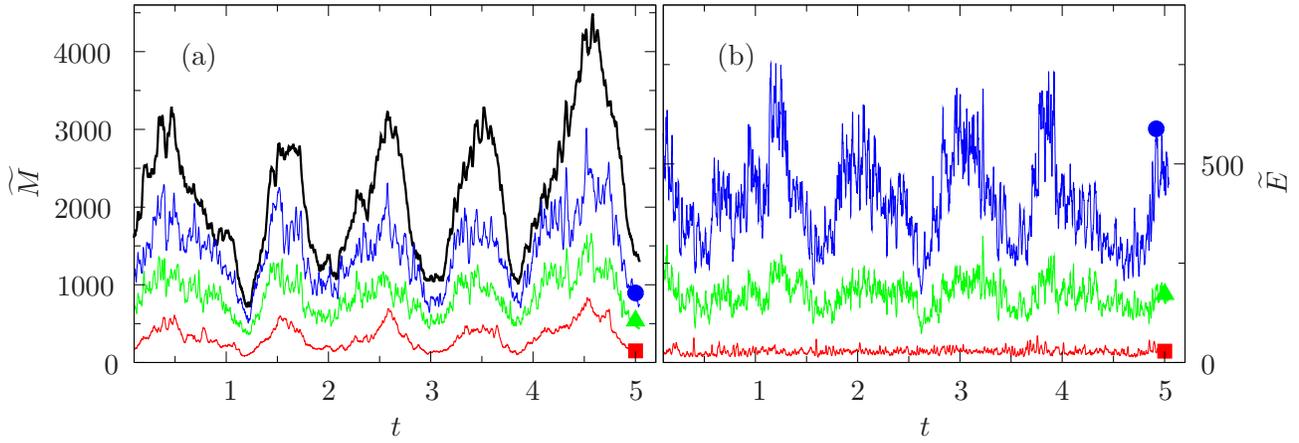}
\end{center}
\caption[]{(Color online)
MD-dynamo dipolar oscillations illustrated by the time series of the
magnetic energy density components (a) and the kinetic energy density
components (b) in the case $P=2$, $\tau=3\times10^4$,
$R=3.5\times10^6$, and $P_m=3$.  
The component $\overline{X}_p$ is shown by thick solid black line,
while $\overline{X}_t$, $\hat X_p$, and 
$\hat X_t$ are 
shown by thin red, green and blue lines, respectively and they
are also indicated by squares, triangles, and circles,
respectively. $X$ stands for either $\widetilde{M}$ or
$\widetilde{E}$.
}
\label{fig-03a}
\end{figure}

\red{In this work, the symbols MD and FD stand for ``mean dipolar''
  and ``fluctuating dipolar'' as they are supposed to indicate the
 dominance of the mean components in the MD case while the fluctuating
 components predominate in the FD dynamos (Simitev and Busse 2009).}
The MD-dynamo is typically realized for Prandtl numbers in excess of
unity (Simitev and Busse 2005). For values of $P$ of the order one it
can be generated if the  
Rayleigh number is not too high. The work of Christensen {\itshape et al.} (1999)
and of Kutzner and Christensen (2002) has exhibited quite well the
transition from MD-type to FD-type dynamos as a function of the
Rayleigh number. As is emphasized in the latter paper, reversals of
the poloidal dipolar field are only found in the narrow transition
region between the two types of dynamos. This result seems to
hold also for the transition region between the two types of dynamos  as a
function of the Prandtl number. 
\red{The transition from FD- to MD-dynamos with increasing P has also
  been observed by Sreenivasan and Jones (2006) who assumed
  $P=P_m$. The property that both types of dynamos can exist at the
  same  parameter values over an extended region of the parameter
  space has been first demonstrated by Simitev and Busse (2009).} 

For the MD-dynamos a description in terms of the magnetostrophic
approximation can be applied since the generation of the geostrophic
zonal flow 
\red{by Reynolds stresses is nearly suppressed. For detailed
  formulations of various versions of this approximation see Simitev
  and Busse (2005). According to the magnetostrophic approximation the
  dependence of dynamos on the Prandtl number disappears. That this
  property is approximately satisfied for MD-dynamos is evident from
  figure \ref{fig-02a}}, 
where the dependence of the components of the kinetic and magnetic
energy densities are plotted as a function of the Prandtl number at
fixed ratio $P_m/P$. 
\red{The various energy density components are defined by
\begin{subequations}
\label{mdens}
\begin{align}
&
\overline{M}_p = \frac{1}{2} \langle \mid \nabla \times ( \nabla \bar h \times \vec r)
\mid^2 \rangle , \quad \overline{M}_t = \frac{1}{2} \langle \mid \nabla \bar g \times
\vec r \mid^2 \rangle, \\
&
\hat{M}_p = \frac{1}{2} \langle \mid \nabla \times ( \nabla \hat h \times \vec r)
\mid^2 \rangle , \quad \hat{M}_t = \frac{1}{2} \langle \mid \nabla \hat g \times
\vec r \mid^2 \rangle,
\end{align}
\end{subequations}
where the angular brackets indicate the average over the fluid shell
and over time and $\bar h$ refers to the azimuthally averaged
component of the poloidal field $h$, while $\hat h$ is defined by
$\hat h = h - \bar h $. Analogous definitions apply for the kinetic
energy densities as given in \eqref{edens} with $E$ replacing $M$.}
Non-geostrophic zonal flows of the thermal wind type are
usually still present in MD-dynamos and are important for the generation of the mean
toroidal field from the axial dipole field. 

On the other hand, a geostrophic zonal flow plays an important role in
the case of 
FD-dynamos and  as a consequence the energies of the mean poloidal and
toroidal magnetic fields are comparable although they are both small
in comparison to the energies of the non-axisymmetric fields. As long
as the Prandtl number is not too low and the Rayleigh number is not
too high, FD-dynamos exhibit oscillations of the axisymmetric
components of the field; see, for example, the paper by Busse and
Simitev (2006). These oscillations assume the form of Parker's (1955)
dynamo waves propagating from the equatorial plane towards higher
latitudes. The oscillations manifest themselves primarily in the mean
toroidal field and sometimes the mean poloidal field participates only
in the form of variations of its amplitude without a reversal of its
sign. Busse and Simitev (2008) have compared this property with the
phenomenon of global excursions of the geomagnetic field which only
sporadically lead to a reversal of the Earth's dipole. 
\begin{figure}
\psfrag{(a)}{(a)}
\psfrag{(b)}{(b)}
\psfrag{(c)}{(c)}
\psfrag{(d)}{(d)}
\psfrag{M/E}{$M/E$}
\psfrag{P}{$P$}
\psfrag{R}{$R$}
\psfrag{O/V}{$O/V$}
\psfrag{M}{$M$}
\psfrag{Nui}{$t$}
\psfrag{0}{0}
\psfrag{1}{1}
\psfrag{10}{10}
\psfrag{20}{20}
\psfrag{2}{2}
\psfrag{3}{3}
\psfrag{4}{4}
\psfrag{5}{5}
\psfrag{6}{6}
\psfrag{7}{7}
\psfrag{8}{8}
\psfrag{x}{$\times$}
\psfrag{Nui}{$Nu_i$}
\psfrag{a4}{$10^4$}
\psfrag{a3}{$10^3$}
\psfrag{a6}{$10^6$}
\begin{center}
\vspace*{3mm}
\epsfig{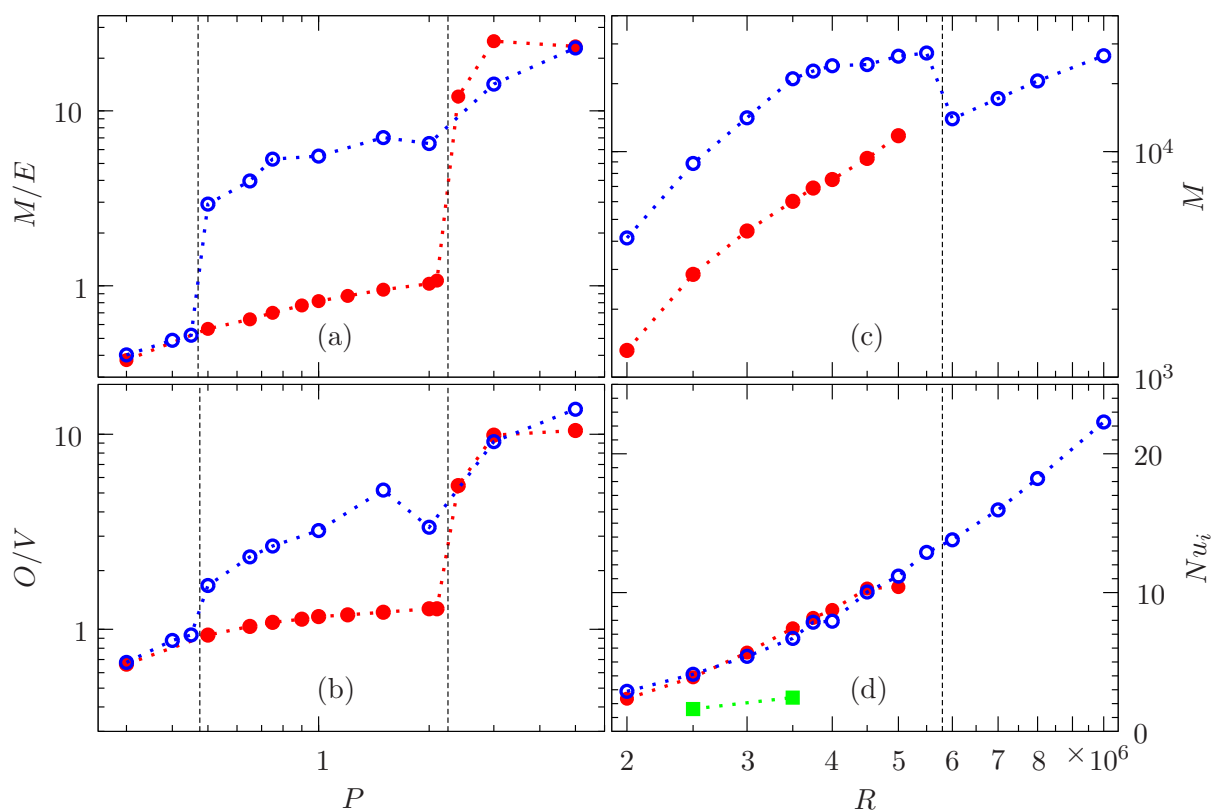}
\end{center}
\caption[]{(Color online) Coexistence and hysteresis of MD and FD
dynamos at $\tau=3\times10^4$.
Panels (a) and (b) show the ratio of the average total magnetic to the
average total kinetic energy, and the ratio of the average ohmic to the
average viscous dissipation, respectively, both as a function of the
Prandtl number in 
the case of $R=3.5\times10^6$, $P/P_m=0.5$.
Panels (c) and (d) show the total magnetic energy density, and the
value of the Nusselt number at $r=r_i$, respectively both as a
function of the Rayleigh number in the case  $P=0.75$, $P_m=1.5$.
Full red and empty blue circles indicate FD and MD dynamos,
respectively. The critical value of $R$ for the onset of thermal
convection for the cases shown in (c) and (d) is $R_c=659145$. 
Values for non-magnetic convection are indicated by green  squares in
(d) for comparison. Vertical dashed lines indicate the approximate
transition values.
}
\label{fig-03}
\end{figure}

MD-dynamos can also exhibit oscillations, but these manifest
themselves just in the amplitude of the various components of the
velocity and the magnetic field. At relatively low Rayleigh numbers
the oscillations are fairly regular with a period of the order of the
viscous time scale as shown in figure \ref{fig-03a}, while at higher
values of $R$ they become more irregular. 
\begin{figure}
\begin{center}
\vspace*{3mm}
\psfrag{E}{$\widetilde{E}$}
\psfrag{M}{$\widetilde{M}$}
\psfrag{t}{$t$}
\psfrag{(a)}{(a)}
\psfrag{(b)}{(b)}
\psfrag{(c)}{(c)}
\psfrag{(d)}{(d)}
\psfrag{0}{0}
\psfrag{2000}{2000}
\psfrag{4000}{4000}
\psfrag{6000}{6000}
\psfrag{8000}{8000}
\psfrag{5000}{5000}
\psfrag{10000}{10000}
\psfrag{1.6}{1.6}
\psfrag{1.8}{1.8}
\psfrag{2}  {2}  
\psfrag{4.2}{4.2}
\psfrag{4.4}{4.4}
\psfrag{4.6}{4.6}
\psfrag{4.8}{4.8}
\hspace*{-8mm}
\epsfig{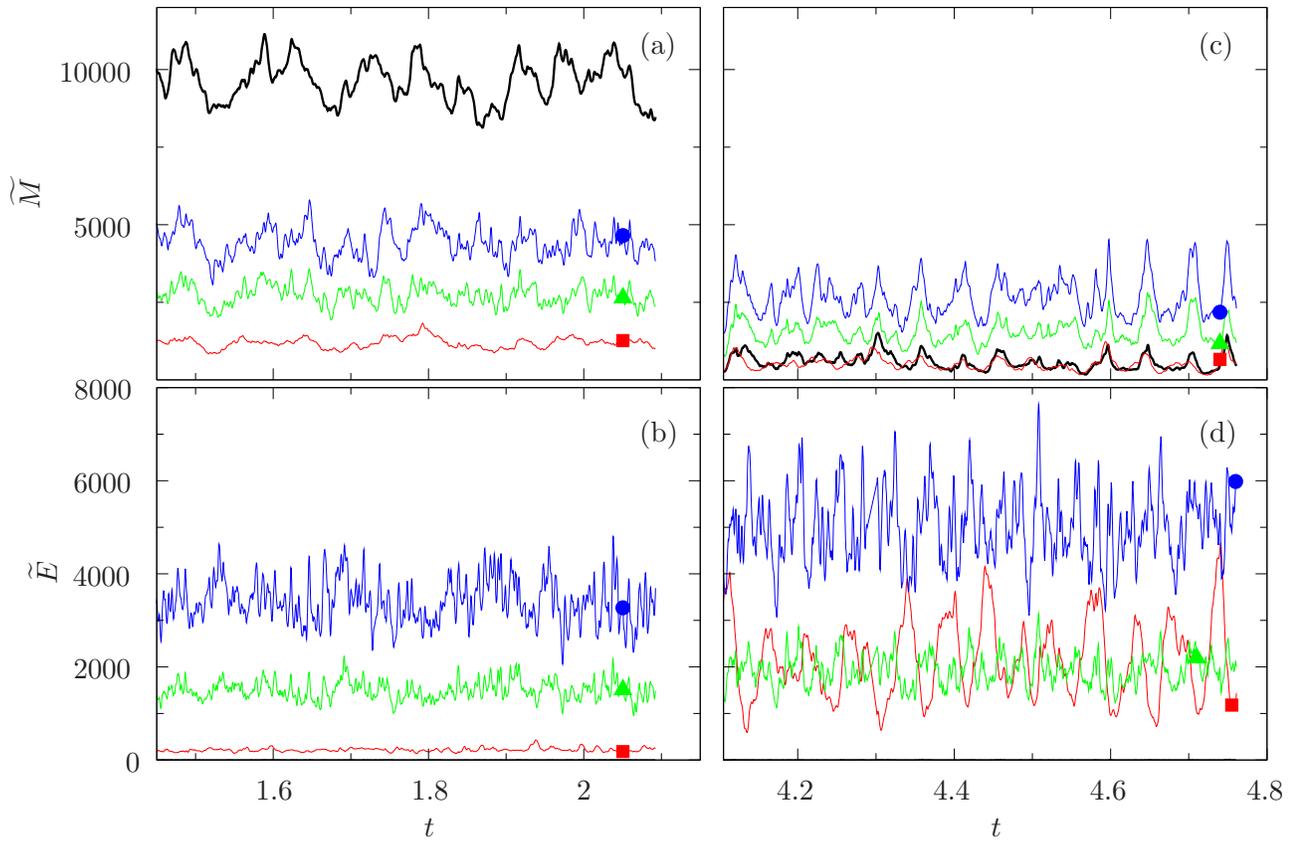}
\end{center}
\caption[]{(Color online)
Time series of two different chaotic attractors are
shown -- a MD (left column (a,b)) and a FD dynamo (right column
(c,d)) both in the case $R=3.5\times10^6$, $\tau=3\times10^4$,
$P=0.75$ and  $P_m=0.75$. The top two panels (a,c) show magnetic energy
densities and the lower two panels show kinetic energy densities. The
component $\overline{X}_p$ is shown by thick solid black line,
while $\overline{X}_t$, $\hat X_p$, and $\hat X_t$ are
shown by thin red, green and blue lines, respectively, and they are
also indicated by squares, triangles, and circles,
respectively. $X$ stands for either $\widetilde{M}$ or $\widetilde{E}$.
}
\label{fig-04}
\end{figure}
 
There is, of course, no sharp boundary between the two types of
dynamos and the dependence of such a not well defined boundary on the
various parameters of the problem including mode of heating has not
yet been investigated in detail. It appears that FD-type dynamos
exist at higher values of $P$ when the rotation rate is lowered.
\red{On the other hand, MD-dynamos tend to exist at lower Prandtl
  numbers when no-slip instead of stress free boundaries are used
  since the Ekman layer tends to damp the geostrophic zonal flow.}
Other types of dynamos may exist in parameter regions distinct from
those that are readily accessible in present computer simulations.  
 
\section{Simultaneous existence of FD- and MD-dynamos}
 
The two types of dynamos can be studied best \red{by direct
 comparison} in the  region of the 
parameter space where they coexist as has been demonstrated in the
paper of Simitev and Busse (2009). Here we wish to discuss further
properties of the coexisting dynamos. In figure \ref{fig-03} the
parameter range of the coexistence of FD- and MD-dynamos has been
indicated in the case $\tau=3\times 10^3$. The average magnetic energy
density $M$ divided by the average kinetic density $E$ and the average ohmic
dissipation divided by the average viscous dissipation, $O/V$, are
shown as function of the Prandtl number $P$ (for fixed $P/P_m$
and fixed $R$) on the left side of the figure. On the right side $M$ and the
Nusselt number are shown as a function of $R$ (for fixed $P$ and fixed
$P_m$). The averages are defined here as the time average of the
spatial average extended over the spherical fluid shell, \red{i.e. $M$ is defined as the time average of $\tilde M=\bar M_p +
\hat M_p + \bar M_t + \hat M_t$. The two types
of dynamos differ also in characteristic ways in the energy densities of
their components.  These energy
densities are highly chaotic as is evident from their time series
shown in figure \ref{fig-04}. Because  the average energy densities in the two
cases are so different, however, even large fluctuations cannot push the
system from one dynamo attractor to the other.}
  
The plot of the Nusselt number $Nu_i$ as function of $R$  in figure \ref{fig-03}
demonstrates that the ability of both dynamos to attain the same
maximal heat transport of convection is the basic reason for their
stability. At the same time the increase of the magnetic energy
density with $R$ indicates that the available buoyancy energy rather
than a constant Elsasser number determines its saturation value. This
finding supports the power laws proposed by Christensen and Aubert
(2006).  
\begin{figure}
\psfrag{mp}{$\overline{p}$}
\psfrag{fp}{$\hat p$}
\psfrag{mt}{$\overline{t}$}
\psfrag{ft}{$\hat t$}
\psfrag{t}{$t$}
\psfrag{mp1}{$\overline{p}_1$}
\psfrag{mp2}{$\overline{p}_2$}
\psfrag{mp3}{$\overline{p}_3$}
\psfrag{fp1}{$\hat p_1$}
\psfrag{fp2}{$\hat p_2$}
\psfrag{fp3}{$\hat p_3$}
\psfrag{mt1}{$\overline{t}_1$}
\psfrag{mt2}{$\overline{t}_2$}
\psfrag{mt3}{$\overline{t}_3$}
\psfrag{ft1}{$\hat t_1$}
\psfrag{ft2}{$\hat t_2$}
\psfrag{ft3}{$\hat t_3$}
\begin{center}
\vspace*{3mm}
\hspace*{-13mm}
\epsfig{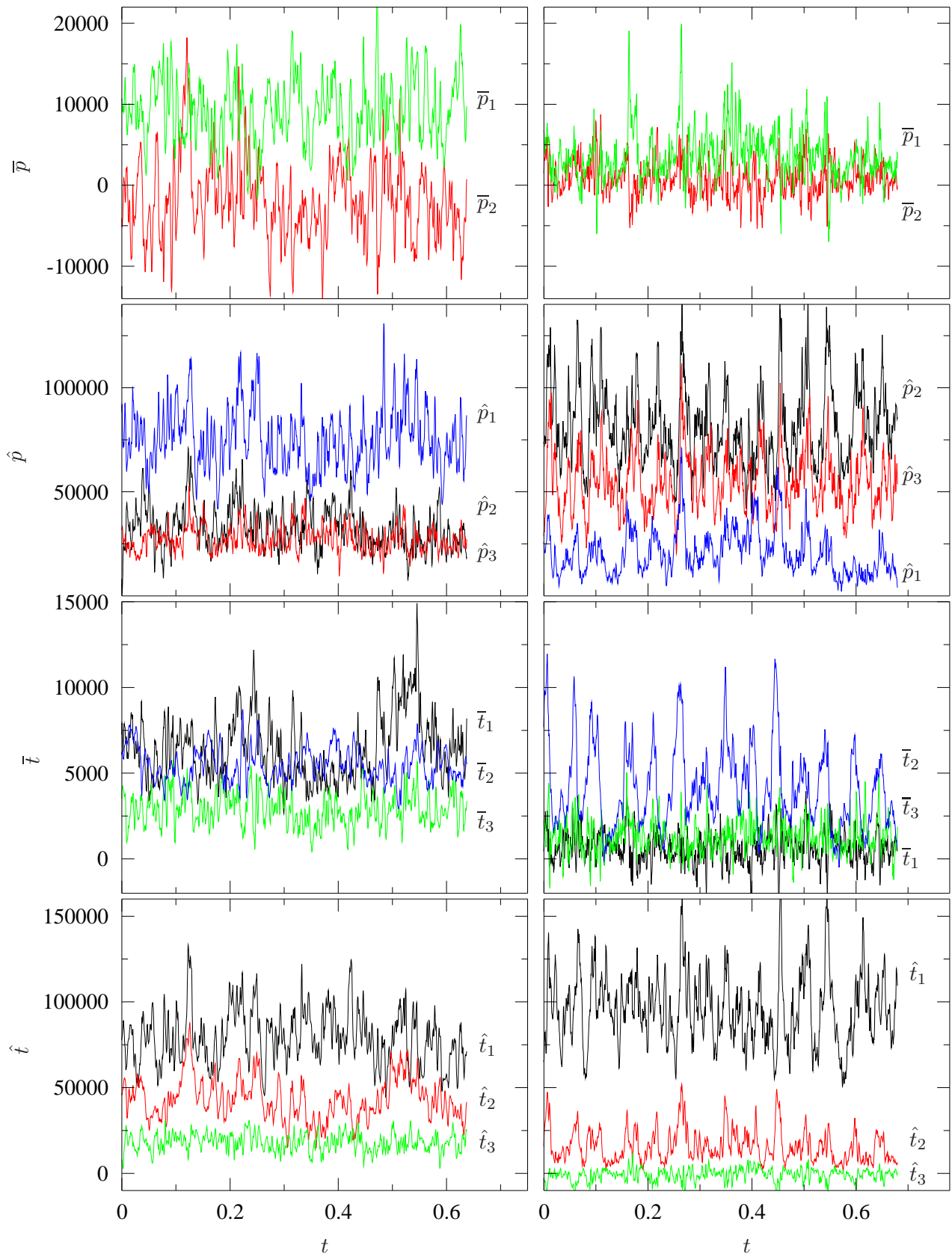}
\end{center}
\caption[]{(Color online)
Comparison of the Lorentz terms of MD (left column) and FD
(right column) dynamos. Both solutions are obtained at identical
parameter values $P=0.75$, $\tau=3\times10^4$, $R=3.5\times10^6$, and
$P_m=1.5$. In the case of the FD dynamo, the mean poloidal
($\overline{p}$), fluctuating poloidal ($\hat p$) and mean toroidal
terms ($\overline{t}$) are multiplied by a factor of 2. The
labels attached to the curves correspond to the notation introduced
in formulae \eqref{lorterms}.
}
\label{fig-05}
\end{figure}
  
It is also of interest to see through which terms the various magnetic
energy densities are generated. In figure \ref{fig-05} the various terms are
shown that are obtained when the $r$-component of the equation of
induction \eqref{1e} is multiplied by $\bar h$ and by $\hat h$ and the
$r$-component of the curl of this equation is multiplied by  $\bar g$
and by $\hat g$ and the results are averaged over the spherical
shell. Following the procedure of Simitev and Busse (2005) we list
here the most important of those terms in the sequence in which they
apply in the case of the MD-dynamo of figure \ref{fig-05},  
\begin{subequations}
\label{lorterms}
\begin{align}
&
\bar p_1 \equiv ( \hat w \hat h \bar h ), \qquad \bar p_2 \equiv
( \hat v \hat g \bar h) ,  \\
&
\hat p_1\equiv( \hat v \bar g \hat h ) ,\qquad \hat p_2\equiv
( \hat v \hat g \hat h ), \qquad \hat p_3\equiv( \hat v \hat h
\hat h), \\
&
\bar t_1 \equiv ( \hat v \hat g\bar g),\qquad \bar t_2 \equiv ( \bar w \bar h \bar g), \qquad \bar t_3 \equiv (\hat w \hat h \bar g),\\
&
\hat t_1\equiv( \hat w \hat g \hat g ) ,\qquad \hat t_2 \equiv( \hat w \bar g \hat g ) ,\qquad  \hat t_3\equiv
( \hat w \bar h \hat g ),
\end{align}
\end{subequations}
where the first two letters inside the brackets indicate which of the 
interactions between velocity and magnetic field components on the right
hand sides of the induction equations  counteract the ohmic dissipation
of the magnetic field component indicated by the last letter inside the
brackets. An alternative interpretation of the averages \eqref{lorterms} is that
they represent the work done by the Lorentz forces on the components
of the velocity field described by $\bar v$, $\hat v$, $\bar w$ and
$\hat w$.  

In comparing the MD-results with the FD-results it is obvious that
axisymmetric MD-components require higher contributions than the
corresponding FD-components \red{as can be seen from the first and third row of figure \ref{fig-05}.
 (Note that FD-results of the first through third row have been
 multiplied by the factor two.) The FD components have smaller
 amplitudes and 
thus suffer less from ohmic dissipation.} On the other hand, the
non-axisymmetric  toroidal field is especially strong and thus require
higher (negative) work done by the Lorentz force to balance its ohmic
losses. \red{Of particular interest are the cases where the order of the amplitude of contributions given by}
\begin{subequations}
\begin{align}
&
\hat p_2\equiv
( \hat v \hat g \hat h ), \qquad \hat p_3\equiv( \hat v \hat h
\hat h), \qquad \hat p_1\equiv( \hat v \bar g \hat h ) ,\\
&
\bar t_2 \equiv (
\hat w \hat h \bar g),\qquad \bar t_3 \equiv ( \hat v \hat g
\bar g), \qquad \bar t_1 \equiv ( \bar w \bar h \bar g).
\end{align}
\end{subequations}
\red{for the FD-dynamos in the case of the second and third row of
  figure \ref{fig-05} differs from the order in the case of
  MD-dynamos.  The axisymmetric components hardly participate in the
  generation of the non-axisymmetric components of the FD-dynamo. On
  the other hand the mean toroidal field of the MD-dynamo is mainly
  amplified by the differential rotation stretching the mean poloidal
  field as suggested by the $\alpha \omega$-dynamo mechanism. The
  latter property contrasts with the results of Olson {\itshape et al.} (1999)
  who find  little evidence for an $\omega$-effect. This difference
  can be explained by the damping of the mean zonal flow by the Ekman
  layer at the  solid outer boundary which is absent in the case of
  the stress-free boundary condition \eqref{vbc} used in the present paper.} 

\section{Concluding remarks}

In this paper we have considered a few assumptions of dynamo theory
which are often taken for granted, but which should be questioned. The
difference between the effects of an imposed magnetic field and one
that is generated by dynamo action is beginning to be appreciated and
an Elsasser number of the order unity is no longer generally accepted
for the determination of the magnetic energy equilibrium. The concept
of eddy diffusivities or of similar methods for representing the
effect of unresolved scales of turbulence will be needed for the
foreseeable future. But the ratios of eddy diffusivities should not
always be set equal to unity. Variations of these ratios can have a
significant influence on convection driven dynamos and may lead to new
insights in the interpretation of observed properties of planetary
dynamos. Finally the common assumption of unique turbulent attractors
should be abandoned and the multiplicity of turbulent states in
nonlinear magnetohydrodynamics  with its huge number of degrees of
freedom should be appreciated. \red{In contrast to some other cases of bistability such as the one observed in
 the VKS-experiment (Berhanu {\itshape et al.} 2009) the states MD and FD do not differ in their symmetry properties.}

\appendix 
\section{.\hspace{3mm}Mathematical formulation of the dynamo problem}

Here a brief outline of the mathematical formulation is given  which
is employed for the simulations of convection driven dynamos referred
to in this paper. It is assumed that a  static state exists with the
temperature distribution $T_S = T_0 - \beta d^2 r^2 /2$ where $d$ is
the thickness of the spherical shell, $rd$ is the length of the
position vector with respect to the center of the sphere and the
gravity field is given by $\vec g = - d \gamma \vec r$. In
addition to the length $d$, the time $d^2 / \nu$,  the temperature
$\nu^2 / \gamma \alpha d^4$ and  the magnetic flux density $\nu (
\mu_0 \varrho )^{1/2} /d$ are used as scales for the dimensionless
description of the problem  where $\nu$ denotes the kinematic
viscosity of the fluid, $\kappa$ its thermal diffusivity, $\varrho$
its density and $\mu_0$ is its magnetic permeability. The equations of
motion for the velocity vector $\vec u$, the heat equation for the
deviation  $\Theta$ from the static temperature distribution, and the
equation of induction for the magnetic flux density $\vec B$ are thus
given by  
\begin{subequations}
\label{eq1}
\begin{gather}
\label{1a}
\partial_t \vec{u} + \vec u \cdot \nabla \vec u + \tau \vec k \times
\vec u = - \nabla \pi +\Theta \vec r + \nabla^2 \vec u + \vec B \cdot
\nabla \vec B, \\
\label{1b}
\nabla \cdot \vec u = 0, \\
\label{1c}
P(\partial_t \Theta + \vec u \cdot \nabla \Theta) = R \vec r \cdot \vec u + \nabla^2 \Theta, \\
\label{1d}
\nabla \cdot \vec B = 0, \\
\label{1e}
\nabla^2 \vec B =  P_m(\partial_t \vec B + \vec u \cdot \nabla \vec B
-  \vec B \cdot \nabla \vec u), 
\end{gather}
\end{subequations}
where $\partial_t$ denotes the partial derivative with respect to time
$t$ and where all terms in the equation of motion that can be written
as gradients have been combined into $ \nabla \pi$. The Boussinesq
approximation has been assumed in that the density $\varrho$ is
regarded as constant except in the gravity term where its temperature
dependence given by $\alpha \equiv - ( \de \varrho/\de T)/\varrho
=${\sl const} is taken into account. The dimensionless parameters obey the
definitions \eqref{params} except for the Rayleigh number $R$ which in the
present spherical case is defined by  
\begin{equation}
R = \frac{\alpha \gamma \beta d^6}{\nu \kappa}. 
\end{equation}
Because the velocity field $\vec u$ as well as the magnetic flux
density $\vec B$ are solenoidal vector fields, the general
representation in terms of poloidal and toroidal components can be
employed,   
\begin{subequations}
\begin{align}
&
\vec u = \nabla \times ( \nabla v \times \vec r) + \nabla w \times 
\vec r \enspace , \\
&
\vec B = \nabla \times  ( \nabla h \times \vec r) + \nabla g \times 
\vec r \enspace .
\end{align}
\end{subequations}
Stress-free boundaries with fixed temperatures are used, 
\begin{align}
\label{vbc}
&
v = \partial^2_{rr}v = \partial_r (w/r) = \Theta = 0 
&
\enspace \mbox{ at }
\enspace r=r_i \equiv 2/3,  
\enspace \mbox{ and at } \enspace r=r_o \equiv 5/3,
\end{align}
where the radius ratio is fixed at the value $r_i/r_o = 0.4$. For the
magnetic field electrically insulating 
boundaries are assumed such that the poloidal function $h$ must be 
matched to the function $h^{(e)}$ which describes the  
potential fields 
outside the fluid shell  
\begin{equation}
\label{mbc}
g = h-h^{(e)} = \partial_r ( h-h^{(e)})=0 \qquad 
\mbox{ at } r=r_i \mbox{ and } r=r_o .
\end{equation}
The energy densities are defined by
\begin{subequations}
\label{edens}
\begin{align}
&
\overline{E}_p = \frac{1}{2} \langle \mid \nabla \times ( \nabla \bar v \times \vec r)
\mid^2 \rangle , \quad \overline{E}_t = \frac{1}{2} \langle \mid \nabla \bar w \times
\vec r \mid^2 \rangle, \\
&
\hat{E}_p = \frac{1}{2} \langle \mid \nabla \times ( \nabla \hat v \times \vec r)
\mid^2 \rangle , \quad \hat{E}_t = \frac{1}{2} \langle \mid \nabla \hat w \times
\vec r \mid^2 \rangle,
\end{align}
\end{subequations}
where the angular brackets indicate the average over the fluid shell  
and $\bar v$ refers to the azimuthally averaged component of $v$,
while $\hat v$ is defined by $\hat v = v - \bar v $. The Nusselt
numbers at the inner and outer spherical boundaries $Nu_i$ and $Nu_o$ are  defined by
\begin{equation}
\label{nu.def}
Nu_i=1- \frac{P}{r_iR} \left.\frac{\de \overline{\overline{\Theta}}}{\de r}\right|_{r=r_i},
 \qquad 
Nu_o=1- \frac{P}{r_oR} \left.\frac{\de \overline{\overline{\Theta}}}{\de r}\right|_{r=r_o},
\end{equation}  
where the double bar indicates the average over the spherical
surface. (The factor $1/R$ has accidentally been dropped in previous
papers of the authors.) The ratio of external heating to internal heating is given by 
\begin{equation}
\frac{r^3_i Nu_i}{r^3_o Nu_o-r^3_i Nu_i}.
\end{equation}

\end{document}